\documentclass[%
 preprint, %linenumbers,
 amsmath,amssymb,
 aps, physrev,
]{revtex4-2}

\usepackage{dcolumn}% Align table columns on decimal point
\usepackage{bm}% bold math
\usepackage{siunitx}
\usepackage{lineno}
\usepackage{caption}
\usepackage{subcaption}
\usepackage[]{hyperref}
\usepackage{braket}
\usepackage{graphicx,import}
\usepackage{tabulary}
\usepackage{tikz}
\usepackage{booktabs}
\usepackage{titlesec}
\newsavebox{\imagebox}
\usepackage{adjustbox}

\begin{document}

%\preprint{APS/123-QED}

\title{\textbf{Multimode Angle‑Compensated Interferometer with Long Time‑Bin Delay Using a Multi‑Pass Herriott Cell} 
}% 

\author{Ramy Tannous}
\email{ramy.tannous@nrc.ca}
\affiliation{National Research Council of Canada, Ottawa, Ontario K1A 0R6, Canada}

\author{St\'ephane Vinet}
\affiliation{Institute for Quantum Computing, University of Waterloo, Waterloo, Ontario N2L 3G1, Canada}
\affiliation{Department of Physics \& Astronomy, University of Waterloo, Waterloo, Ontario N2L 3G1, Canada}
%\altaffiliation[Current address: ]{Quantum Labs, Cisco Systems, 3232 Nebraska Ave, Santa Monica, California 90404, USA}

\author{Kaylee Sherk}
\affiliation{Institute for Quantum Computing, University of Waterloo, Waterloo, Ontario N2L 3G1, Canada}
\affiliation{Department of Physics \& Astronomy, University of Waterloo, Waterloo, Ontario N2L 3G1, Canada}

\author{Kimia Mohammadi}
\affiliation{Institute for Quantum Computing, University of Waterloo, Waterloo, Ontario N2L 3G1, Canada}
\affiliation{Department of Physics \& Astronomy, University of Waterloo, Waterloo, Ontario N2L 3G1, Canada}

\author{Thomas Jennewein}
\affiliation{Institute for Quantum Computing, University of Waterloo, Waterloo, Ontario N2L 3G1, Canada}
\affiliation{Department of Physics \& Astronomy, University of Waterloo, Waterloo, Ontario N2L 3G1, Canada}
\affiliation{Department of Physics, Simon Fraser University, Burnaby, British Columbia V5A 1S6, Canada}

%\collaboration{CLEO Collaboration}%\noaffiliation

\date{\today}% It is always \today, today,
             %  but any date may be explicitly specified

\begin{abstract}
Interference of optical signals in free-space channels requires optical receivers to support many spatial modes due to atmospheric turbulence, typically necessitating adaptive optics systems. Angle-compensated interferometers offer a passive alternative, making them particularly attractive for time-bin encoded signals with delays on the order of one nanosecond. Here, we demonstrate a multimode angle‑compensated interferometer design that achieves a high interference visibility for spatially multimode beams with large time bin separations. The interference of the multimode beams is enabled using an imaging multi-pass Herriott cell that enables a very long path separation with a small form-factor. The design is tested using both numerical ray-tracing simulations and proof-of-principle demonstrations. We create a prototype interferometer with a path length difference of 12ns and determine that it maintains a high interference visibility with a large field-of-view of \SI{0.4}{\degree}.
%Development of field-deployable interferometers are vital in the development of a global quantum network and sensing applications. 
\end{abstract}

\maketitle

%%%%%%%%%%%%%%%%%%%%%%%%%%  body  %%%%%%%%%%%%%%%%%%%%%%%%%%
%%%%%%%%%%%%%%%%%%%%%%%%%%  body  %%%%%%%%%%%%%%%%%%%%%%%%%%
\section{Introduction}

%Quantum networking and communication is gaining popularity as many governments and institutions invest in its development and deployment. 
Time-bin encoding has been the preferred method for encoding quantum information for use in optical fiber networks for decades. Time-bin encoding has also been demonstrated to be feasible over free-space channels \cite{jin2019genuine}, and more recently over highly multimode optical fibers which are commonly used in server and data centers~\cite{chen2018multimode,bhatt2025continued,tannous2025towards}, making them practical to use for quantum interconnects. However, interfacing with stationary qubits, such as solid-state devices, trapped ions, or cold atomic ensembles, typically requires temporal separations sufficient to resolve individual time bins and to match the absorption bandwidths and storage dynamics of these systems \cite{clausen_quantum_2011,saglamyurek_broadband_2011, saha_high-fidelity_2025}. While such long delays can be readily implemented using fiber delay lines, realizing them with free-space interferometers is significantly more challenging.
%Achieving low error rates over a turbulent free-space and highly multimode optical fibers can be challenging due to the loss of interference visibility associate with trying to interfere spatially distorted beam in an unbalanced interferometer.
Spatial mode distortions arising from propagation through multi-mode optical channels compromise photon indistinguishability within unbalanced interferometric architectures, resulting in decreased interference visibility and significant error rates. This phenomenon is particularly pronounced in mobile links, where telescope pointing error and atmospheric turbulence lead to dynamic fluctuations of the interferometer's optical path length difference \cite{jin2018demonstration,vallone2016interference}. Adaptive optics offer a promising solution by compensating for atmospheric distortions in real-time thus enabling efficient coupling into single-mode fibers. However, adaptive optics systems can be lossy and require additional overhead that may not be desired for certain platforms, e.g. nano-satellites. An alternative approach that has been used extensively by the optical community is to create an angle‑compensated or field-widened interferometer~\cite{Hilliard:66,hirschberg_field_1974,vallone2016interference,jin2018demonstration,jin2019genuine,cahall_multi-mode_2020}. A field-widened interferometer is a special case of an angle‑compensated interferometer that increases the solid angle accepted by an interferometer, $\theta$. This can be achieved by carefully selecting and designing the optical path of each interferometer arm by use of optically dense material~\cite{cahall_multi-mode_2020,sajeed_observing_2021} or an imaging system~\cite{vallone2016interference,jin2018demonstration,tannous2026all}.
\par Another important challenge for free-space implementations is the physical footprint of the interferometer. Achieving long time-bin delays requires large path length differences, leading to increasingly bulky and impractical setups. For multimode angle‑compensated interferometers, the use of optically dense materials is challenging as the time delays get large. Stabilizing interferometers can also be difficult as the path length increases. For passive temperature stabilization, it is typically desired to use a folded optical path to reduce the insulation needs. Although it has been shown that one can still achieve good interference over multi-mode channels with a standard interferometer~\cite{tretiakov2024multi}, the performance drops as the time delay increases and requires precise alignment of the system. To overcome these limitations, compact multi-pass geometries such as Herriott cells offer an attractive solution \cite{Herriott:65}, as they allow long optical path lengths to be folded into a small footprint while maintaining alignment stability. In this work, we demonstrate a field-widened interferometer leveraging a Herriott cell design to handle spatially multimode beams and long time delays within a compact form factor. %Adaptive optics can allow for coupling into single mode fiber and undo any distortions present. %However, adaptive optics systems can be lossy and require additional overhead that may not be desired for certain platforms, e.g. nano-satellites.

\section{Designs}
\begin{figure}[htbp]
    \centering
    \includegraphics[width=0.5\linewidth]{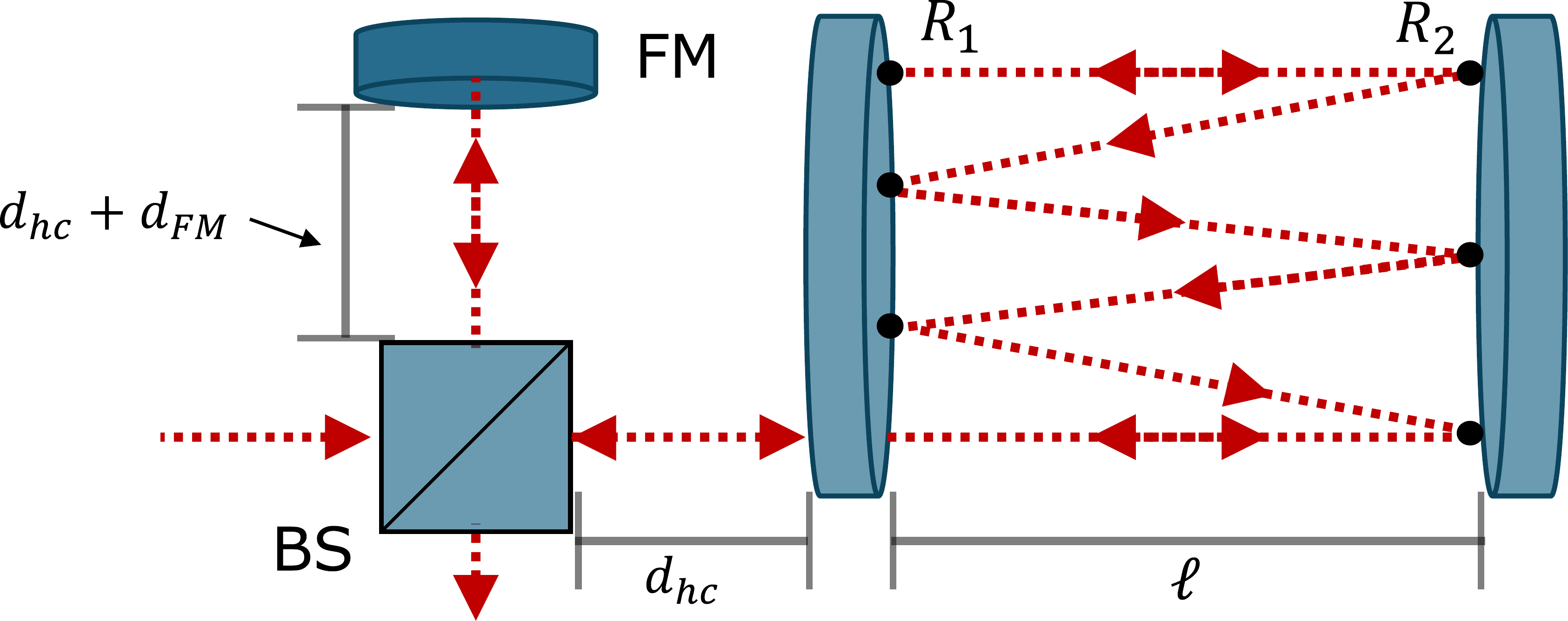}
    \caption{Conceptual design of the Herriott cell interferometer.}
    \label{fig:conceptual_design}
\end{figure}

 The conceptual design of a Herriott cell interferometer is shown in Fig.~\ref{fig:conceptual_design}, illustrated here in a Michelson configuration but readily adaptable to a Mach–Zehnder layout. The critical design parameters are the radius of curvature of the two spherical mirrors of the cell $R_1$ and $R_2$, and the separation between the two mirrors, $\ell$.

 For this study, we limit the Herriott cell patterns to the horizontal plane, however all the analysis and discussion can apply to Herriot cell patterns that span both the horizontal and vertical planes, allowing for longer path delay. Further investigation such patterns and the placement of the aperture is beyond the scope of this work.
 
\subsection{Matrix Method}
To determine the conditions of a multimode angle‑compensated interferometer, the parameters of the Herriott cell are tuned such that the output of each path of the Michelson overlaps regardless of the angle of incidence. We use a ray transfer matrix method for accurate non-sequential 3D ray tracing to determine the position ($x$,$y$,$z$) and directional cosine ($s_x$,$s_y$,$s_z$) of the output ray of each path. Thus the ray transfer matrices for various optical elements can be described as a $6\times6$ matrix. For example a simple ray propagation is given by,
\begin{equation}
    \label{eq:3d_prop}
    \begin{bmatrix}
        x'\\
        y'\\
        z'\\
        s_x'\\
        s_y'\\
        s_z'\\
    \end{bmatrix}
    =
    \begin{bmatrix}
        1 & 0 & 0 & d & 0 & 0\\
        0 & 1 & 0 & 0 & d & 0\\
        0 & 0 & 1 & 0 & 0 & d\\
        0 & 0 & 0 & 1 & 0 & 0\\
        0 & 0 & 0 & 0 & 1 & 0\\
        0 & 0 & 0 & 0 & 0 & 1\\
    \end{bmatrix}
    \begin{bmatrix}
        x\\
        y\\
        z\\
        s_x\\
        s_y\\
        s_z\\
        
    \end{bmatrix},
\end{equation}
where $d$ is the length of the ray and is usually easily determined using geometry. Other simple $6\times6$ ray transfer matrices for simple optical actions can be easily derived e.g. reflection from a flat mirror. For the more complicated Herriott cell transfer matrix, we follow the analysis of Ref.~\cite{cao2020modified}. The ray transfer matrix is derived by determining the coordinate of intersection of the ray with the surface of the spherical mirror, as well as the direction of the ray relative to the normal of the mirror surface~\cite{cao2020modified}. By applying propagation and vector reflection theory, the coordinate of the $n+1$ reflected ray in the Herriott cell pattern can be determined from the $n^\text{th}$ ray by,

\begin{equation}
    \label{eq:3d_hc_prop}
    \begin{bmatrix}
        x^{n+1}\\
        y^{n+1}\\
        z^{n+1}\\
        s_x^{n+1}\\
        s_y^{n+1}\\
        s_z^{n+1}\\
    \end{bmatrix}
    =
    \begin{bmatrix}
        1 & 0 & 0 & d_n & 0 & 0\\
        0 & 1 & 0 & 0 & d_n & 0\\
        0 & 0 & 1 & 0 & 0 & d_n\\
        0 & 0 & 0 & 1-\frac{2F_x^2}{F_x^2+F_y^2+F_z^2} & -\frac{2F_xF_y}{F_x^2+F_y^2+F_z^2} & -\frac{2F_xF_z}{F_x^2+F_y^2+F_z^2}\\
        0 & 0 & 0 & -\frac{2F_yF_x}{F_x^2+F_y^2+F_z^2} & 1-\frac{2F_y^2}{F_x^2+F_y^2+F_z^2} & -\frac{2F_yF_z}{F_x^2+F_y^2+F_z^2}\\
        0 & 0 & 0 & -\frac{2F_zFx}{F_x^2+F_y^2+F_z^2} & -\frac{2F_zF_y}{F_x^2+F_y^2+F_z^2} & 1-\frac{2F_x^2}{F_z^2+F_y^2+F_z^2}\\
    \end{bmatrix}
    \begin{bmatrix}
        x^n\\
        y^n\\
        z^n\\
        s_x^n\\
        s_y^n\\
        s_z^n\\
        
    \end{bmatrix},
\end{equation}
where $d_n$ is the length of the $n^\text{th}$ ray and $F_x=\partial F/\partial x$, $F_y=\partial F/\partial y$, $F_z=\partial F/\partial z$ are the partial derivatives of the mirror surface equation. The mirror surface equation is given by,
\begin{equation}
    \label{eq:mirror_equation}
    F(x,y,z)=x^2+y^2+2(-1)^{n+1}(z-z_0)R_{n+1}+(z-z_0)^2=0,
\end{equation}
where $z_0=[1+(-1)^{n+1}]\ell/2$, and $R_{n+1}$ is determined by the value of $n+1$,

\begin{equation}
    \label{eq:R}
    R_{n+1}=\begin{cases}
        R_1 & n+1 \text{ odd}\\
        R_2 & n+1 \text{ even}.
    \end{cases}
\end{equation}

To determine the ray propagation of each $n^\text{th}$ step, we use numerical methods to solve Eq.~\ref{eq:3d_hc_prop} for points that satisfy the mirror equation, Eq.~\ref{eq:mirror_equation}. From this we can get a complete transfer matrix of each path of the interferometer and can determine the resulting ray's spatial coordinates and directional cosines. Each $n+1$ ray is calculated by first determining the $\mathbf{x_{n+1}}$ from the $\mathbf{x_n}$ and $\mathbf{s_n}$ by solving the equation $\mathbf{x_{n+1}}=\mathbf{x_n}+\mathbf{s_n}\cdot\mathbf{d}$ (taken from Eq.~\ref{eq:3d_hc_prop}) for $d$ for points $x,~y,~z =x^{n+1},~y^{n+1},~z^{n+1}$ that satisfy $F(x^{n+1},~y^{n+1},~z^{n+1})=0$. Once $\mathbf{x_{n+1}}$ is determined, we use the lower half of Eq.~\ref{eq:3d_hc_prop} with the condition of $F(x^{n+1},~y^{n+1},~z^{n+1})=0$ to determine $\mathbf{s_{n+1}}$. 

\begin{figure}[htbp]
    \centering
    \begin{subfigure}{0.495\columnwidth}
        \includegraphics[width=\textwidth]{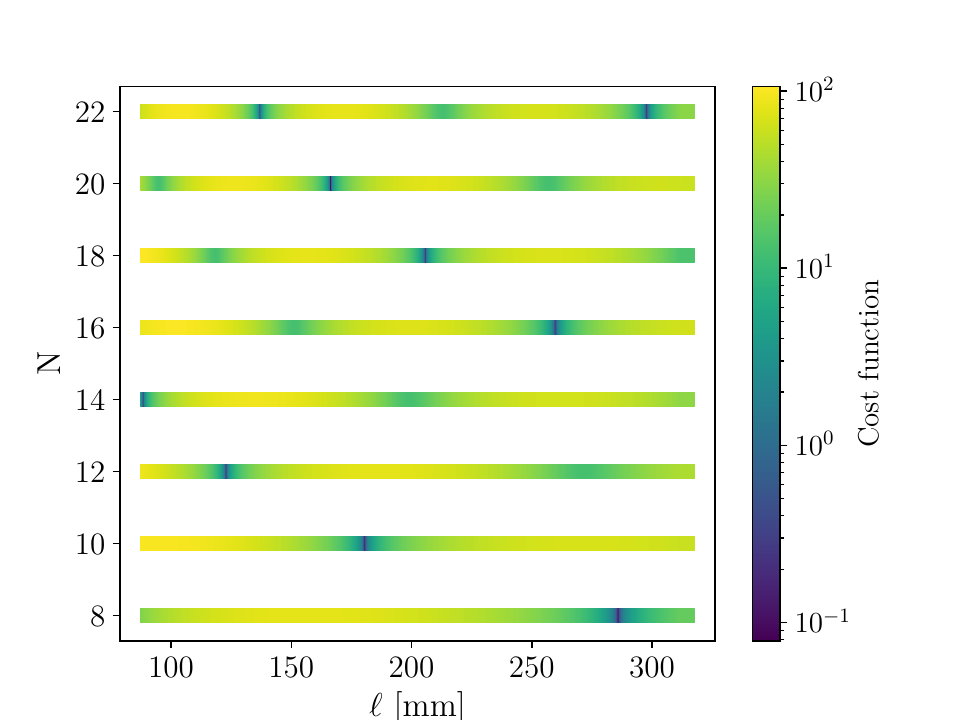}
        \caption{}
        \label{fig:analytical_N_L}
    \end{subfigure}
    \begin{subfigure}{0.495\linewidth}
        \includegraphics[width=1\linewidth]{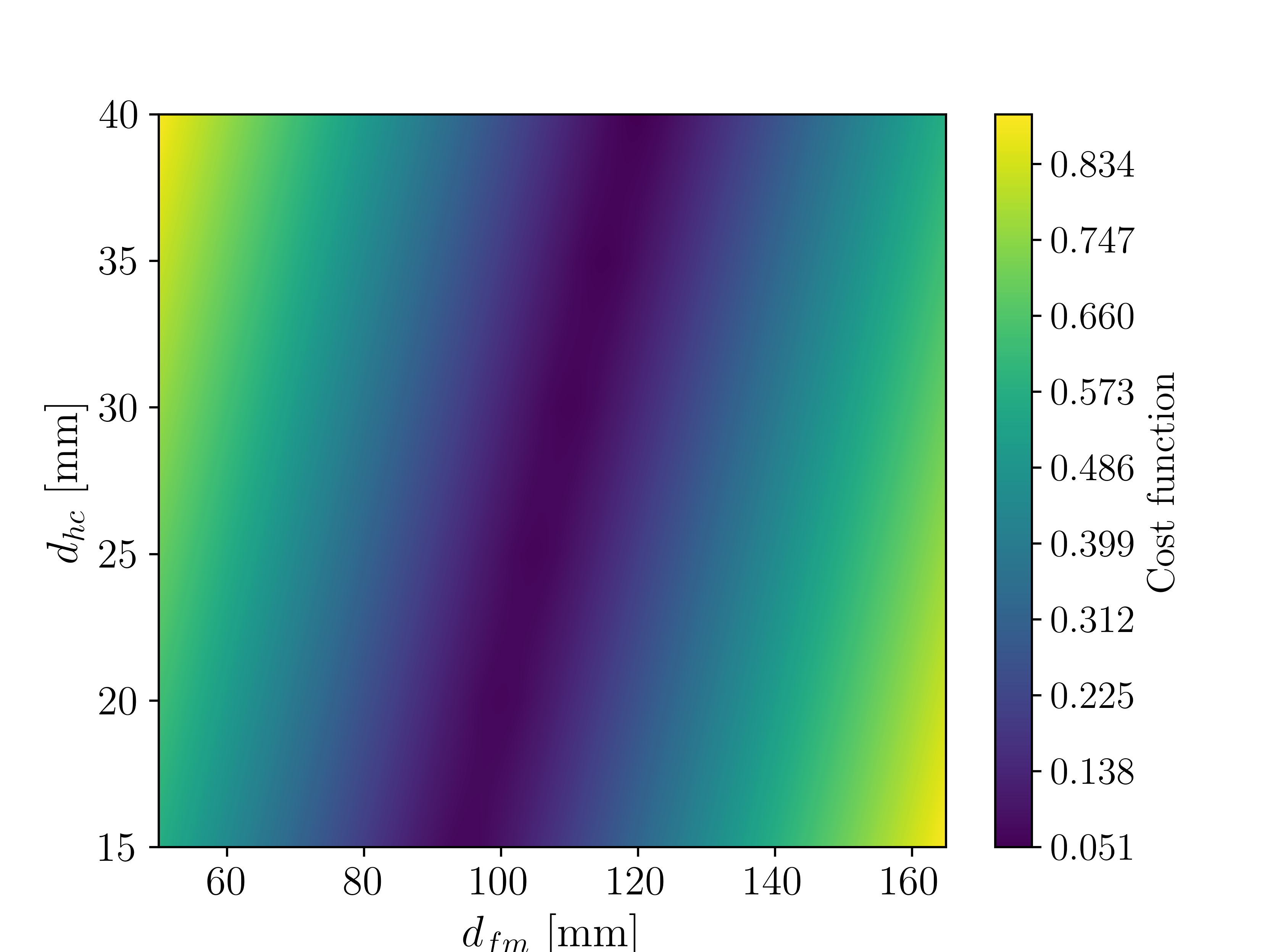}
        \caption{}
        \label{fig:analytical_df_dhc}
    \end{subfigure}
    \caption{(\subref{fig:analytical_N_L}) Analytical optimization of the HC to be an imaging system, the colour bar represents the norm-1 distance cost function.  (\subref{fig:analytical_df_dhc}) Analytical optimization of the distance $d_{hc}$ and $d_{fm}$ for $N=18$ with $\ell$ at its optimal value. For both plots $R_1=$~\SI{1000}{\milli\meter} and $R_1=$~\SI{700}{\milli\meter}.}
    \label{fig:analytical_analysis}
\end{figure}

We then compare the output rays of the Herriott cell ($N^\text{th}$ ray) and the flat mirror path using the norm-1 distance as a cost function and optimize $\ell$ and $d_{fm}$ to minimize the cost function. This ensures the proper overlap of the output beam for both the spatial coordinates $\mathbf{x}$ and the directional cosines $\mathbf{s}$. This optimization ensures that the beams are overlapped at the output, however it does not flatten the angular sensitivity of the optical path delay (OPD), the defining condition of a field-widened interferometer. Instead the optimization simply minimizes the angular sensitivity of the OPD, future studies will investigate configurations that also satisfy the field-widened condition. In this work $d_{hc}$, $R_1$, and $R_2$ are fixed; however, these parameters could be included as free parameters in the optimization. Furthermore, while we do not fix the time-bin separation of the interferometer in our optimization, it could be easily set as a fixed design requirement. Fig.~\ref{fig:analytical_analysis} shows the results of the matrix analytical analysis for $R_1=$~\SI{1000}{\milli\meter} and $R_1=$~\SI{700}{\milli\meter}. Fig.~\ref{fig:analytical_N_L} shows the optimization of the values of $\ell$ for various Herriott cell bounces ($N$). The optimization of $\ell$ is done to ensure that $s^{N}_{x,y}=s^0_{x,y}$ and $s^{N}_{z}=-s^0_{z}$, which is similar to a flat mirror. Once $\ell$ is determined, the optimization of $d_{fm}$ is done to ensure spatial overlap of the beams of the two optical paths, i.e. $\mathbf{x^N}=\mathbf{x^0}$. Fig.~\ref{fig:analytical_df_dhc} shows the norm-1 distance between $\mathbf{x^N}$ and $\mathbf{x^0}$ for various $d_{fm}$ and $d_{hc}$ combinations with $N=18$ and $\ell=$~\SI{208.63}{\milli\meter}, the optimal value determined in Fig.~\ref{fig:analytical_N_L}. Although the 3D ray tracing method is an accurate way to simulate and describe the optical system, wavefront distortions, polarization, and phase effects are not captured using this method. Furthermore, our analytical analysis can yield several values of $\ell$ that satisfy the overlap conditions, however not all are physical due to the geometric limits of the Herriott.

\subsection{Gausslet Ray Tracing}
\label{sec:GaussletRayTracing}
We verify the 3D ray tracing matrix method analysis using a non-sequential ray tracing package that includes wavefront distortions and polarization~\cite{Cole_2021}. The package uses Gaussian Beamlet Decomposition (GBD) as the main method of propagating a wavefront~\cite{harvey2015modelingGausslet,ashcraft2020openGausslet}. GBD is done by propagating a superposition of Gaussian beams as complex rays. A wavefront is decomposed into a finite set of Gaussian beamlets (Gausslets). The Gausslets are then propagated and then coherently added to recreate the resulting wavefront. Each Gausslet is fully parameterized by a central ray that tracks the position of the Gausslet, and a complex curvature matrix that describes its waist radius and curvature. Both the central ray and complex curvature matrix are propagated using geometrical ray tracing. GBD is a fast technique that is capable of modeling optical systems for diffraction, interference, tilt/decenter errors, and polarization. 

 As seen in Fig.~\ref{fig:analytical_analysis}, there are several solutions where the angle‑compensated condition is satisfied. However, not every solution for $\ell$ is physically achievable or realistic with standard 2-inch optics. Some realistic considerations not captured by the matrix method are; the aperture size of the Herriott cell entrance aperture and the finite width of the spherical mirrors. Therefore, the non-sequential GBD ray tracing is used to verify the matrix method calculations for realistic optical designs. %Fig.~\ref{fig:raypier_hc} shows an image of the GBD ray tracing.
 
\begin{figure}[htbp]
    \centering
    % \begin{subfigure}{0.495\linewidth}
    %     \includegraphics[width=1\columnwidth]{figures/place_holders_nb/raypier_hc.png}
    %     \caption{}
    %     \label{fig:raypier_hc}
    % \end{subfigure}
    % \\
    %\begin{subfigure}{0.495\linewidth}
        \includegraphics[width=1\columnwidth,trim={5cm 0 5cm 0}, clip]{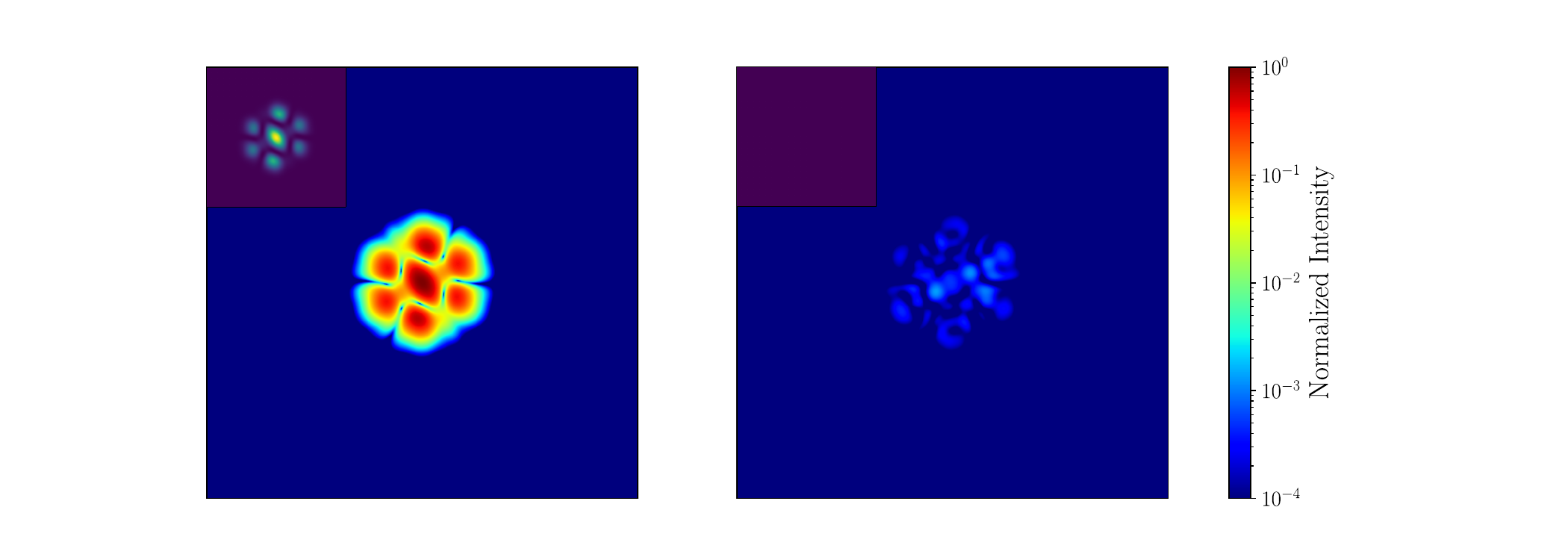}
        %\caption{}
        %\label{fig:raypier_spot}
    %\end{subfigure}
    
    \caption{(%\subref{fig:raypier_hc}) Ray tracing of the the HC interferometer. (\subref{fig:raypier_spot}) 
    Ray tracing simulation, constructive and destructive interference intensities in logarithmic scaling. Insets show the intensity in linear scaling.}
    \label{fig:raypier_spot}
\end{figure}

 Given that GBD simulations are capable of modeling interference and polarization effects, the performance of various HC interferometer designs are verified. We quantify the performance by the interference visibility of the modeled device. The visibility is determined by $V=\frac{I_{max}-I_{min}}{I_{max}+I_{min}}$, where $V=1$ is the maximum visibility and $I_{max},~I_{min}$ are the output intensities for constructive and destructive interference ($\pi$ phase difference) respectively. For the HC interferometer to be considered a multimode angle-compensated interferometer, it must be able to maintain a high visibility for both a single-mode Gaussian input wavefront and a highly structure multi-mode wavefront (Fig.~\ref{fig:raypier_spot}). In addition, a high visibility should be maintained for different angles of incidence, $\theta$. Thus in the GBD simulations, the visibility is investigated for different input wavefronts and the angle of incidence. The fringe visibility is achieved by slightly varying the flat mirror position (short path) of the HC interferometer such that the output signal experiences constructive and destructive interference.

\begin{figure}[htbp]
    \centering
    \begin{subfigure}{0.495\linewidth}
        \includegraphics[width=\linewidth]{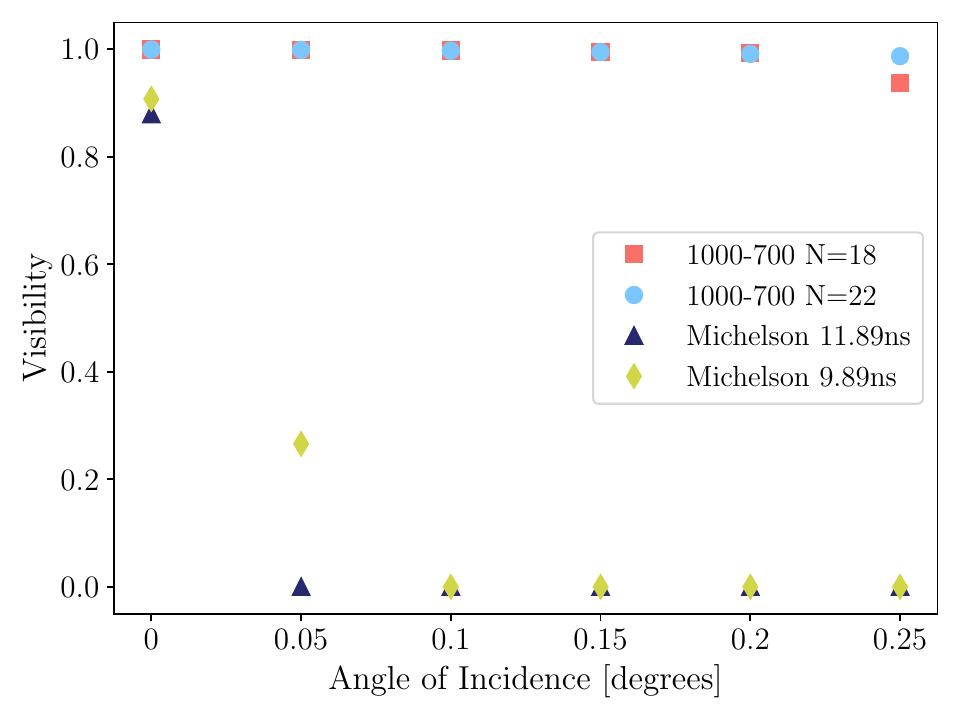}
        \caption{Gaussian single mode input}
        \label{fig:aoi_sm_sim}    
    \end{subfigure}
    \begin{subfigure}{0.495\linewidth}
            \includegraphics[width=\linewidth]{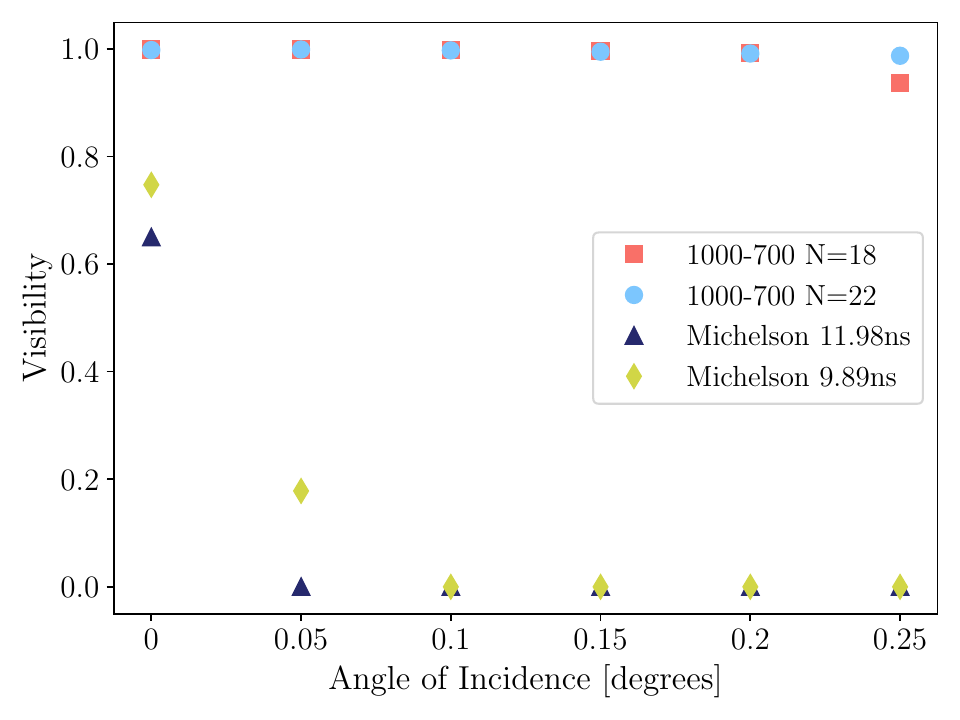}
        \caption{Multi-mode input}
        \label{fig:aoi_mm_sim}
    \end{subfigure} 
    
    \caption{Ray tracing results of a varying angle of incidence test with a Gaussian single mode (\subref{fig:aoi_sm_sim}) and a multi-mode (\subref{fig:aoi_mm_sim}) input beam. The for uncorrected Michelson interferometers with similar path delays are shown for comparison.}
    \label{fig:raypier_aoi}
\end{figure}

 As shown in Fig.~\ref{fig:raypier_spot}, the HC interferometer produces high quality destructive interference, even for a highly multi-mode beam. Furthermore, the high interference visibility is maintained for various angles of incidence of the input beam. The results of the simulations for some select interferometer designs are shown in Table~\ref{tab:designs}. The single-mode Gaussian input is lower-bounded by the spatially multi-mode input visibility (MM vis) and thus not reported in Table~\ref{tab:designs}. The visibility as a function of angle of incidence is shown in Fig.~\ref{fig:raypier_aoi}. The visibility remains high for the HC interferometers and is not reduced by the change in angle of incidence or multi-mode input. An uncorrected Michelson interferometer of the same path delay suffers significantly from small changes in the angle of incidence or changes to the input mode spatial structure. The sudden drop in the interference visibility for the HC interferometer at a specific angle of incidence is due to clipping at the entrance aperture of the HC. The clipping can occur at the input or during any of the Herriott cell round trips, depending on the spot pattern of the HC.

% add how many designs existed, state we chose fixed radius of curvature but this could be a degree of freedom.
Table~\ref{tab:designs} shows the details of some select interferometer designs for interferometers with a path difference of $\geq$\SI{10}{\nano\second}. An important factor in the Herriott cell design is the distance of the entrance aperture from the center of the spherical mirror. For our purposes we restrict the placement this aperture at a fixed distance of \SI{20}{\centi\meter} for 2-inch optics and \SI{7}{\centi\meter} for 1-inch optics.  Although in this work $R_1$, $R_2$ and $d_{hc}$ are kept constant, they can be included as free parameters in the design optimization, especially if a given interferometer time delay is required.

\begin{table}[htbp]
\centering
\caption{Herriott cell interferometer design. All units are in \si{\milli\meter} unless specified otherwise. The visibilities are determined using the GBD simulations. The bold-face entry is implemented and tested, the discussion is in Sec.~\ref{sec:exp}.}
\label{tab:designs}
\begin{tabular}{@{}llllllll@{}}
\toprule
$R_1$ & $R_2$ & $\ell$ & $d_{fm}$ & N & $d_{hc}$ & $\Delta t$ [\si{\nano\second}] & MM vis \\ \midrule
 \textbf{1000}  &  \textbf{700}  &  \textbf{208.63 }  &  \textbf{ 99.79 } & \textbf{18}  &  \textbf{20}   & \textbf{ 11.98} & \textbf{0.997}  \\
 1000 &  700  &   139.93  &  64.93   &  22 &   20  &  9.89 & 0.996 \\
 1000 &  700  &  359.35   &  144.28   & 20  &   20  &  23.13 & 0.997 \\
 1000 &  600  &  187.23   &   82.74  & 18  &   20  &  10.81 & 0.997  \\
 700 &  700  &  324.78   &  181.26   & 26  &   20  & 27.07 & 0.996  \\
 700 &  600  &  273.89   &  138.41   &  14 &   20  & 11.99 & 0.995  \\
 700  & 500   &  241.03   &  106.02   &  14 &  20   & 10.67 & 0.997  \\ \bottomrule
\end{tabular}%

\end{table}

%%%%%%%%%%%%%%%%%%%%%%%%%%%%%%
% Experimental Tests, time bin and vis
%%%%%%%%%%%%%%%%%%%%%%%%%%%%%% 

\section{Prototype and Testing}
\label{sec:exp}
\begin{figure}[htbp]
    \centering
        \begin{subfigure}{0.48\linewidth}
        \centering
         \includegraphics[clip,trim={0.5 cm} {1 cm} {0.5 cm} {1.83 cm},width=\linewidth]{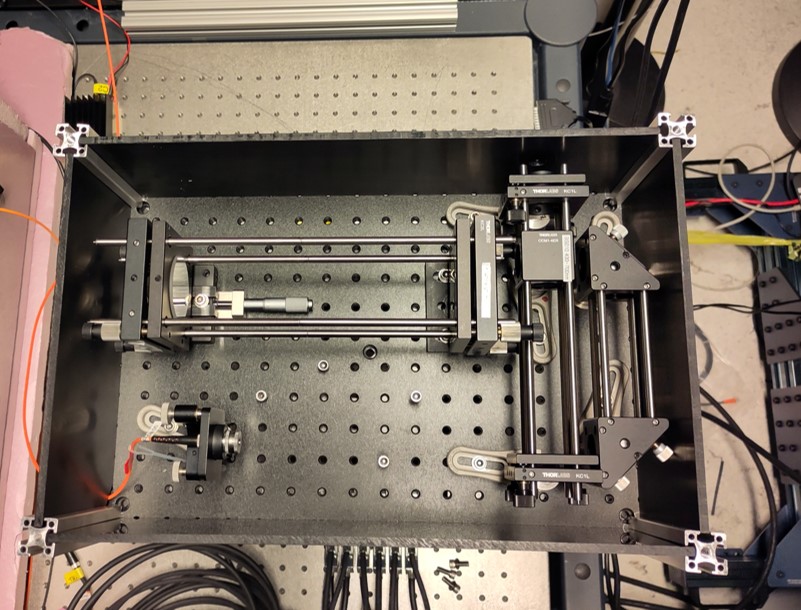}
         \caption{Top view}
         \label{fig:aoi_top}    
     \end{subfigure}
    \begin{subfigure}{0.48\linewidth}
        \centering
        \includegraphics[clip,trim={1 cm} {3 cm} {0.75 cm} {2.99 cm},width=\linewidth]{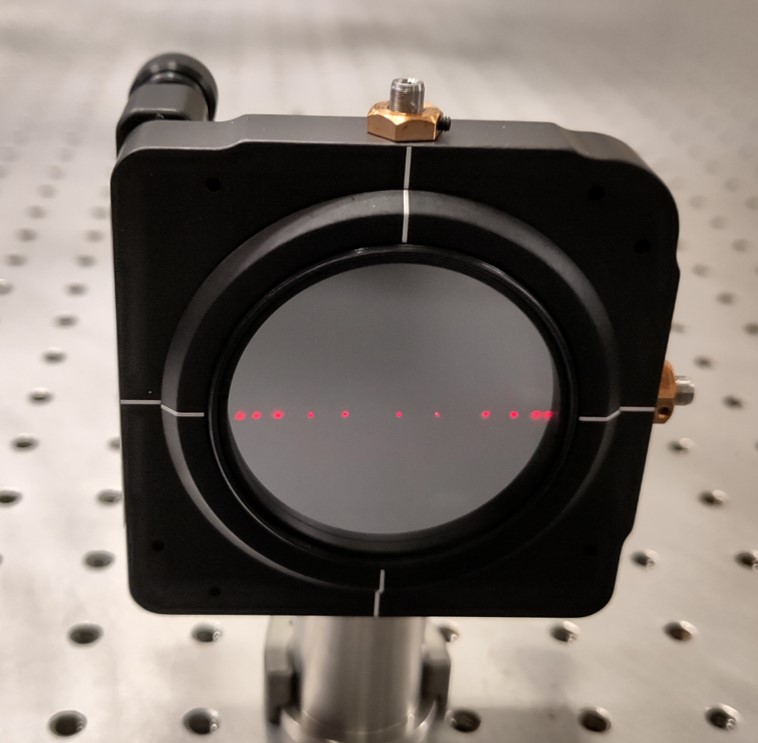}
        \caption{}
        \label{fig:spot_pattern_real}    
    \end{subfigure}
    \caption{~(\subref{fig:aoi_top})~Top view of the HC interferometer setup.~(\subref{fig:spot_pattern_real})~Horizontally spaced spots forming the HC pattern on the spherical mirror.}
    \label{fig:setup_hcint}
\end{figure}

A prototype of the first entry (bold-face) of Table.~\ref{tab:designs} is built using standard bulk optics and optomechanics. The mirrors of the Herriott cell mirrors are 2-inch optics. The entrance mirror were custom ordered and had a \SI{3}{\milli\meter} hole \SI{20}{\milli\meter} from the center. The interferometer is built in a Michelson configuration similar to Fig.~\ref{fig:conceptual_design} and is shown in Fig.~\ref{fig:aoi_top}. One of the interferometer outputs/inputs that has a fiber launcher that is connected to a multi-mode fiber (OM4), the other is left as a free-space input/output for testing purposes. Fig.~\ref{fig:spot_pattern_real} shows the horizontal spot pattern on one of the HC mirrors. For the purposes of this proof-of-concept demonstration, we do not stabilize the phase of the interferometer. Although the enclosure does provide some phase stability, long term stability of the interferometer would require the use of phase stabilization techniques that are well studied \cite{ma2012experimental,roztocki2021arbitrary,vsvarc2023sub}.

\begin{figure}[htbp]
    \centering
    \begin{subfigure}{0.35\linewidth}   
        \centering
        \includegraphics[width=\linewidth]{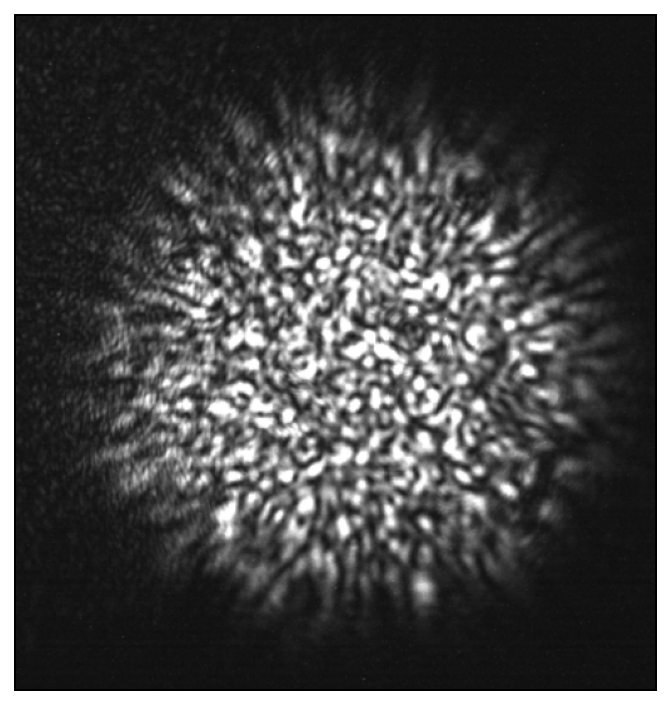}
        \caption{}
        \label{fig:cw_mmf_spot}
    \end{subfigure}
    \begin{subfigure}{0.48\linewidth}
        \centering
        \includegraphics[width=\linewidth]{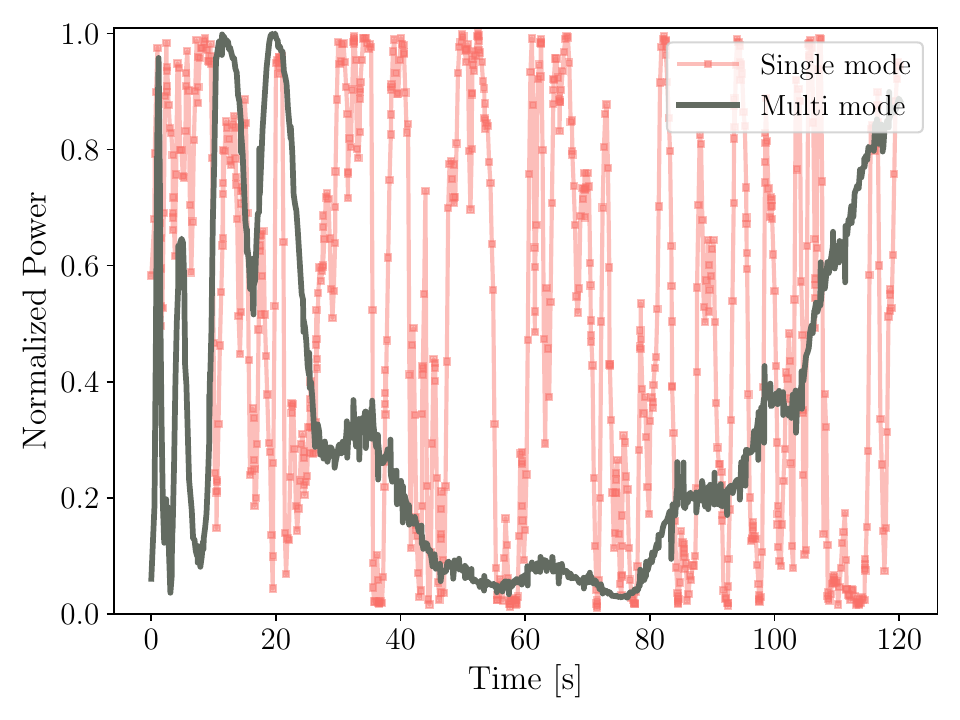}
        \caption{}
        \label{fig:cw_mmf_res}
    \end{subfigure}
    \caption{(\subref{fig:cw_mmf_spot})~Highly multimode beam used for visibility tests with a long coherence continuous-wave laser.~(\subref{fig:cw_mmf_res})~Experimental visibility data with the \SI{785}{\nano\meter} continuous-wave laser for a single and multi mode input. The power is measured using a Thorlabs PM100A powermeter. The interferometer phase is not stabilized.}
    \label{fig:cw_mmf_vis}
\end{figure}

Once build the interferometer is tested using long coherent continuous wave (cw) lasers. We use two different lasers, a CrystaLaser (CL532-200-S) at \SI{532}{\nano\meter} and a Toptica (DL-pro) at \SI{785}{\nano\meter}. We use two different lasers in the visible and near infrared to demonstrate the practicality of the interferometer as it can be used with a broad range of wavelengths. Both signals were tested using a single mode Gaussian, and a spatially multi-mode input. The visibility for the single mode Gaussian input and spatially multi-mode for each wavelength are shown in Table.~\ref{tab:res}. The visibility is measured using a power meter and by allowing the relative phase of the two interferometer paths to drift. The discrepancy between the results for the visibility with the \SI{532}{\nano\meter} and \SI{785}{\nano\meter} is explained by the differences in the reflectance of the HC mirror coatings. The mirrors are coated with a silver coating that has a $>98\%$ reflectance at normal incidence. An ideal implementation would use a dielectric coating that has a very high reflectance for the wavelength range of choice. Fig.~\ref{fig:cw_mmf_spot} shows the structure of the spatially multi-mode beam at the output of the interferometer.

\begin{table}[htbp]
    \centering
    \caption{Visibility results for the various scenarios the HC interferometer is tested. Note that there is no change to the optical components. }
    \label{tab:res}
    \begin{tabular}{@{}llll@{}}
    \toprule
     Input mode & CrystaLaser (\SI{532}{\nano\meter}) & Toptica DL-Pro (\SI{785}{\nano\meter}) & Simulation \\ \midrule
     Gaussian single mode & 0.991(1) & 0.980(1) & 0.998  \\
     Multi-mode & 0.970(2) & 0.951(2) &  0.997 \\
     \bottomrule
\end{tabular}%

\end{table}

%%%%%%%%%%%%%%%%%%%%%%%%%%%%%%
% Experimental Tests, AOI
%%%%%%%%%%%%%%%%%%%%%%%%%%%%%%

An aspect of the HC interferometer is to maintain high visibility despite the input angle of incidence. That is, the interference visibility should be invariant with changes in the input beam angle. We confirm that the HC interferometer can achieve this using the GBD ray tracing and experimental confirmation. In the ray tracing, the angle of the input beam is varied until the visibility drops. The drop in visibility is due to beam-clipping on the entrance aperture of the Herriott cell either at the input or throughout the optical path inside the cell.

\begin{figure}[htbp]
    \centering
    \begin{subfigure}{0.495\linewidth}
        \centering
        \includegraphics[clip,trim={0.5 cm} {0 cm} {2.5 cm} {2.5 cm},width=\linewidth]{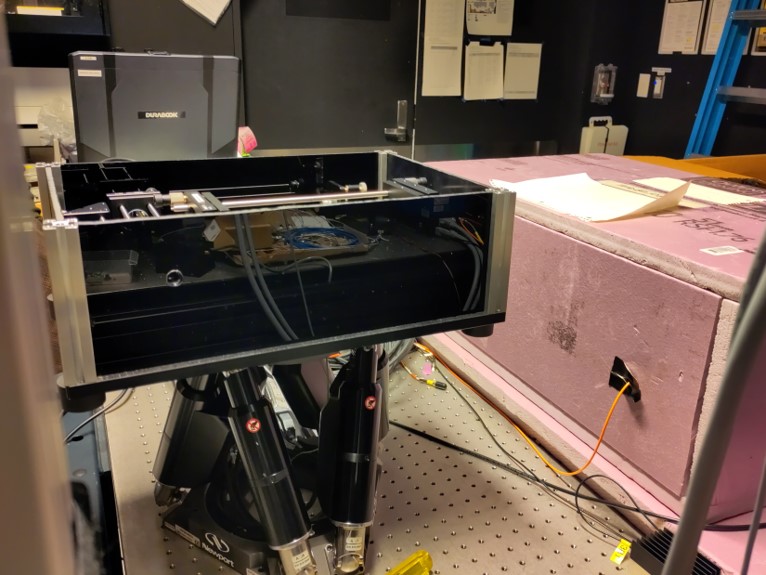}
        \caption{Front view}
        \label{fig:aoi_font}    
    \end{subfigure}
    % \begin{subfigure}{0.489\linewidth}
    %     \includegraphics[width=\linewidth]{figures/place_holders_nb/aoi_test_top.jpg}
    %     \caption{Top view}
    %     \label{fig:aoi_top}    
    % \end{subfigure}
    % \\
    \begin{subfigure}{0.495\linewidth}
         \includegraphics[width=\linewidth]{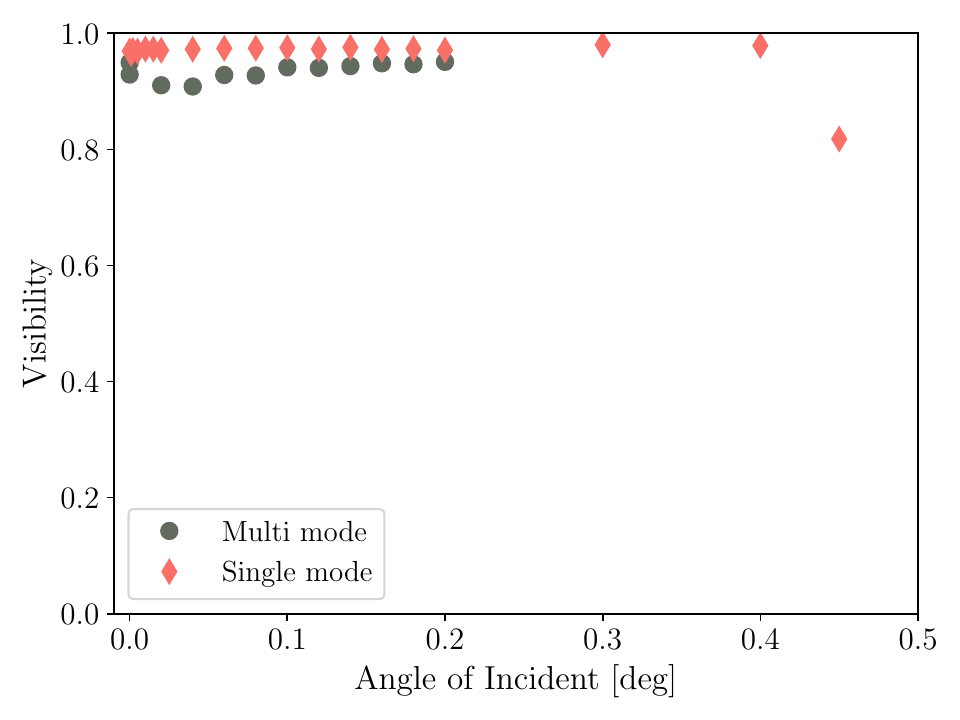}
         \caption{ }
        \label{fig:aoi_res}
    \end{subfigure}
    \caption{ ( \subref{fig:aoi_font}) The angle-of-incidence setup with the interferometer on the six-leg hexapod (Newport). %(\subref{fig:aoi_top}) Top view of the experimental setup.
    (\subref{fig:aoi_res}) The experimental angle of incident results for single and multi-mode input beams.}
    \label{fig:aoi_ex_setup}
\end{figure}

\begin{figure}[htbp]
    \centering
    \begin{subfigure}{0.496\linewidth}
        \begingroup
           \sbox0{\includegraphics{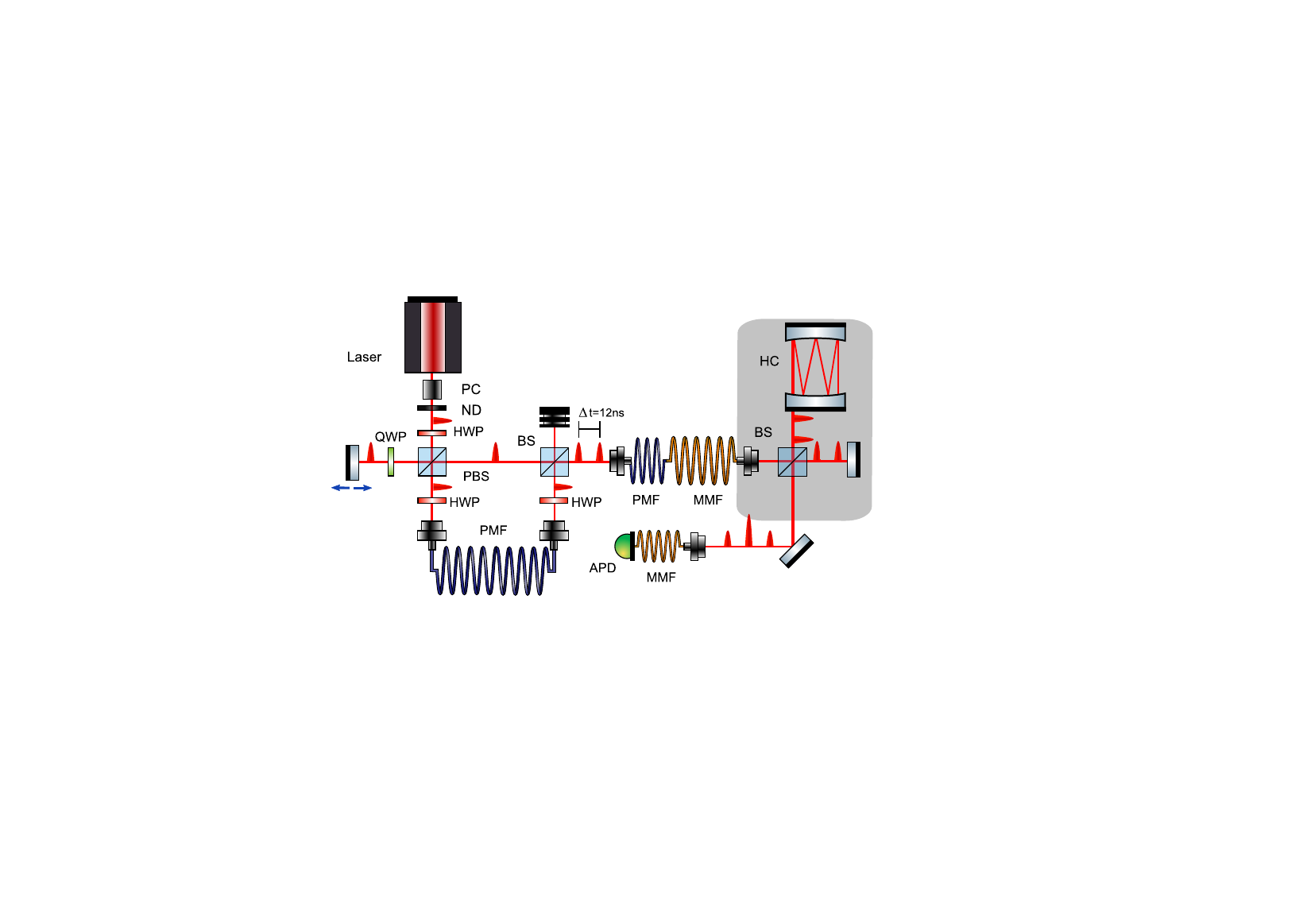}}%
           \includegraphics[clip,trim={0.25\wd0} {0.25\wd0} {0.33\wd0} {0.225\wd0},width=\linewidth]{figures/tb_fiber_setup.pdf}
            \caption{Experimental Setup}
            \label{fig:tb_ex_setup}
        \endgroup
    \end{subfigure}
    \begin{subfigure}{0.495\linewidth}
            \includegraphics[width=\linewidth]{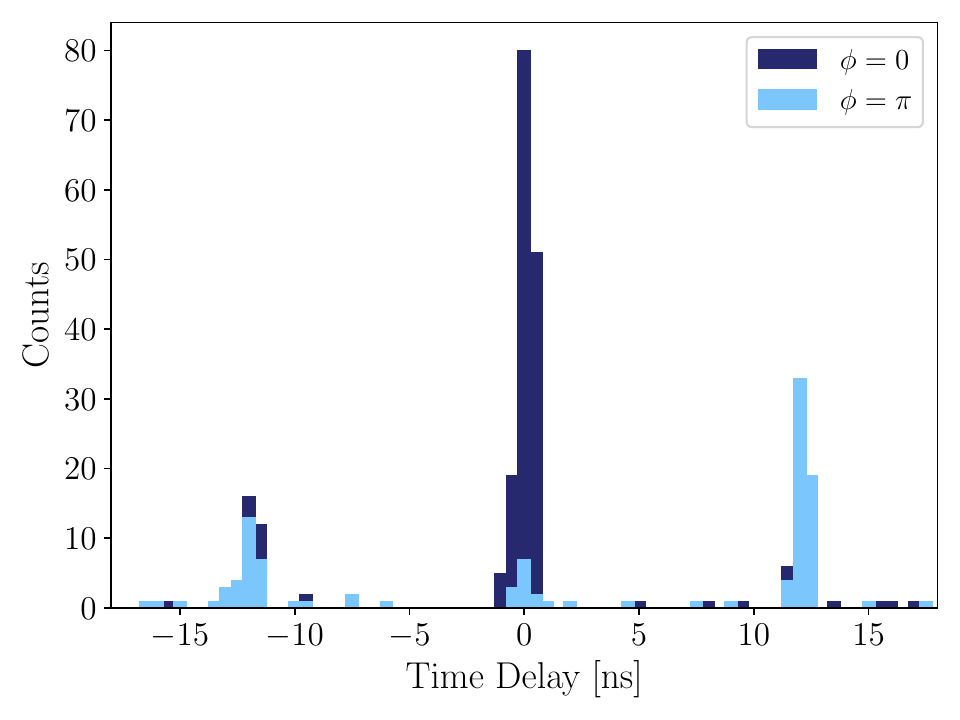}
        \caption{Time-Bin Histogram}
        \label{fig:tb_min_max}
    \end{subfigure}    
    \caption{ The time-bin experimental setup (\subref{fig:tb_ex_setup}). An unbalanced fiber interferometer is used to create an early and late time bin. A \SI{5}{\meter} long multi-mode fiber is used to distort the pulses spatial mode. The HC interferometer interferes the time bins, and an APD is used to measure the pulses to create a histogram. (\subref{fig:tb_min_max}) Histogram of the time bins after passing through both interferometers. The time-bin separation is \SI{12}{\nano\second}. The darker data is for constructive interference ($\phi=0$) while the lighter data is for destructive interference ($\phi=\pi$).}
    \label{fig:tb_experiment}
\end{figure}

% Time bin separation plots
Finally, the interferometer is tested using an attenuated pulsed laser. The time-bin separation is shown to match the expected \SI{12}{\nano\second} of the design. Figure.~\ref{fig:tb_ex_setup} shows the experimental setup while figure~\ref{fig:tb_min_max} shows the resulting histograms for the photon time of arrival after passing through both interferometers. The \SI{12}{\nano\second} time-bins are created using an attenuated \SI{1}{\mega\hertz} pulsed laser at \SI{785}{\nano\meter} passing through an unbalanced fiber interferometer. The pulses then travel across a \SI{5}{\meter} multi-mode fiber (OM4) such that the spatial mode is no longer single-mode, and then are sent through the HC interferometer where the superposition time-bin visibility is measured. The time histogram of the output of the HC interferometer is shown in figure~\ref{fig:tb_min_max}. The dark histogram bar data is for when the relative path phase difference is $\phi=0$ (constructive interference), while the light histogram bar data is for when the relative path phase difference between the early and later time bins is $\phi=\pi$ (destructive interference). The superposition basis interference visibility is measured to be $0.88$. The relative drop in visibility compared to the CW experiments could be due to an imbalance in the beamsplitters, the higher loss of the long path in both the fiber-based and HC interferometer, and a mismatch in the time delay between the two interferometers. Furthermore, dispersion in the fiber-based interferometer could contribute to some drop in visibility though the amount should be minimal due to the \SI{10}{\pico\second} pulses used. Nonetheless, the visibility is quite high considering the large time delay and the highly multi-mode channel, and is sufficient for quantum key distribution and entanglement swapping experiments \cite{sun2017entanglement,saha_high-fidelity_2025}.

%%%%%%%%%%%%%%%%%%%%%%%%%%%%%%%%%%
\section{Summary and Conclusion}
%%%%%%%%%%%%%%%%%%%%%%%%%%%%%%%%%%

In summary, we present an optical architecture for a field-widened time-bin interferometer design that uses multi-pass cells. The design employs reflective optics for the imaging system, enabling compatibility with multiple single-photon sources across different wavelength ranges and facilitating spectral multiplexing. %The interferometer design uses reflective optics for the imaging system which allows for direct use with multiple single photon sources at different wavelength ranges, and for spectral multiplexing.
We experimentally demonstrated that the reflective optical design performs well as an interferometer with a multimode interference visibility of greater than $0.9$ for a time delay of \SI{>10}{\nano\second}.  Furthermore, the folded optical paths Herriott cell produces long relative path delays with a significantly reduced form factor, allowing for very long optical path length differences compared to a standard field-widened Michelson approach. Although Herriott cells have been used in interferometers for gravitational wave detection~\cite{weiss1972electromagnetically,hough_search_2005,adhikari_gravitational_2014}, this is the first known usage which in a multimode angle-compensated interferometer.
Importantly, despite the large optical path difference, the interferometer preserved high visibility even at increased input beam angles, highlighting its robustness to angular deviations. This tolerance is especially relevant for free-space quantum communication, where pointing errors and atmospheric turbulence induce beam wander and angle-of-incidence variations that often degrade interferometric stability. 
%The interferometer also maintained a high visibility despite an increase in the angle-of-incidence of the input beam, which is especially impressive for an interferometer with a large optical path difference. 
%In fact, this work presents the longest interferometer path difference for a field-widened interferometer.

Although our investigation in this manuscript is limited to the configurations shown in Table.~\ref{tab:designs}, different spherical mirror radii and optic sizes can be used to increase or decrease the relative path delay. Future work is to investigate satisfying the field-widened condition and the limits of the design in terms of optical relative path delay and form factor, for example, long time delays approaching the hundreds of nanoseconds regime for delay line memory applications that could be used with spatial mode encoding~\cite{10.1117/12.3003228,guo2025highlyefficientbroadbandoptical,guo2026high}. In addition, the investigation and integration of shorter delay designs (i.e. $\approx1$~\si{\nano\second}) is of interest, as these time delays are useful for most high rate quantum networking systems. Beyond time-bin encoding, field-widened interferometers have recently been employed for frequency-bin quantum communication \cite{Vinet:25, vinet_time-resolved_2026}. The Herriott-cell approach developed here could extend these capabilities by reducing the footprint of frequency-bin analyzers and enabling high-dimensional decoding via cascaded interferometers. Furthermore, owing to its inherent compatibility with both temporal and spectral modes, this design could facilitate the implementation of hybrid time–frequency communication protocols. The significantly reduced form factor is advantageous for field deployment in resource-limited applications such as satellites, and may enable alternative integration strategies, including additive manufacturing of the optical housing%for using alternative manufacturing techniques to store the optical components such as additive manufacturing~
\cite{TannousRamy2023}. Finally, our analysis in this work was limited to Herriott cell patterns in a single plane, longer delays may be achievable with utilizing all three physical dimensions, however, optical alignment complexity may increase.

\begin{acknowledgments}
R.T. would like to thank Philip Bustard and Benjamin Sussman for helpful discussions. This project was funded by the National Research Council of Canada New Beginnings Project Ideation Fund. S.V. acknowledges support from the Natural Sciences and Engineering Research Council of Canada (NSERC) through a Canada Graduate Scholarship--Doctoral (CGS-D).
\end{acknowledgments}

\section*{Conflict of Interest}
The authors declare no conflicts of interest.

\section*{Data Availability}
The data underlying the results presented in this paper are not publicly available at this time but may be obtained from the authors upon reasonable request.

\bibliography{apssamp}% Produces the bibliography via BibTeX.

\end{document}